\documentclass{article}
\usepackage{epsfig}
\begin{document}
{\hfill{FTUV-00-0601}}
\vspace{2cm}

\begin{center}
{\Large{\bf{Chiral Unitary Approach to the $\bar{K}$ Nucleus Interaction
and Kaonic Atoms}}}
\end{center}
\vspace{1.5cm}

\begin{center}
\Large{ E. Oset$^1$, S.Hirenzaki$^2$, Y. Okumura$^2$, A. Ramos$^3$, 
H. Toki$^4$ and M.J. Vicente Vacas$^1$}
\end{center}
\begin{center}
{\small\it{$^1$Departamento de F\'{\i}sica Te\'orica and IFIC,
Centro Mixto Universidad de Valencia-CSIC, Valencia, Spain. \\
$^2$Physics Department, Nara Women University, Nara, Japan.\\
$^3$Departament d'Estructura i Constituents de la Materia, Universitat
de Barcelona, Spain\\
$^4$RCNP, Osaka University, Osaka, Japan}}
\end{center}
\vspace{1cm}
\begin{abstract}
{\small{ We review recent work on various topics related to the
modification of kaon properties in nuclei. After a brief exposition of the
$\bar{K}N$ and $\bar{K}$ nucleus interaction, results from the application to
$K^-$ atoms, renormalization of the $f_0$ and $a_0$ scalar resonances in nuclei
and $\phi$ decay in the nucleus are shown. }}
\end{abstract}

\section{ Introduction}

   In this talk I shall be reporting on the properties of antikaons in matter
with applications to kaonic atoms, the renormalization of the $f_0(980)$ and
$a_0(980)$ scalar meson resonances in nuclei and the $\phi$ decay in nuclei.
The starting point is the $\bar{K} N$ interaction which we study using a
coupled channel unitary method by means of chiral Lagrangians. The medium
modifications are included in the $\bar{K}N$ amplitude and an integration is
done over the Fermi sea of occupied nucleons. This selfenergy is then used in
the Klein Gordon equation to find out bound states of the kaons in nuclei,
which appear in two families, the atomic states and deeply bound nuclear
states. The same selfenergy is  used in the chiral unitary coupled 
channel approach which leads to the $f_0(980)$ and $a_0(980)$ resonances in 
order to investigate the modification of these resonances in nuclei. Finally, 
the decay modes of the $\phi$ in nuclei and its total width are studied.
 
\section{The properties of the $\bar{K}$ in the nuclear medium}

 In this section we address the properties of the $\bar{K}$ in the 
nuclear medium which have been studied in  \cite{knuc}. The work is based on
the elementary $\bar{K} N$ interaction which was obtained in \cite{oset98}
using a coupled channel unitary approach with chiral Lagrangians. 
  
The lowest order chiral Lagrangian, coupling the octet of pseudoscalar 
mesons to the octet of $1/2^+$ baryons, is
\begin{eqnarray}
L_1^{(B)} = &&\langle \bar{B} i \gamma^{\mu} \nabla_{\mu} B
\rangle -
M_B \langle \bar{B} B \rangle \nonumber \\
& + &\frac{1}{2} D \langle \bar{B} \gamma^{\mu} \gamma_5 \left\{
u_{\mu},
B \right\} \rangle
+ \frac{1}{2} F \langle \bar{B} \gamma^{\mu} \gamma_5 [u_{\mu},
B]
\rangle \ ,
\end{eqnarray}
where the symbol $\langle\, \rangle$ denotes the trace of SU(3)
matrices. Expressions for the different magnitudes can be found in \cite{oset98}.

The coupled channel formalism requires to evaluate the transition
amplitudes between the different channels that can be built from
the meson and baryon octets. For $K^- p$ scattering there are 10 such
channels, namely $K^-p$, $\bar{K}^0 n$, $\pi^0
\Lambda$, $\pi^0 \Sigma^0$,
$\pi^+ \Sigma^-$, $\pi^- \Sigma^+$, $\eta \Lambda$, $\eta
\Sigma^0$,
$K^+ \Xi^-$ and $K^0 \Xi^0$. In the case of $K^- n$ scattering
the coupled channels are: $K^-n$, $\pi^0\Sigma^-$,
 $\pi^- \Sigma^0$, $\pi^- \Lambda$, $\eta
\Sigma^-$ and
$K^0 \Xi^-$.

 At low energies the transition amplitudes can be written as 
\begin{equation}
V_{i j} = - C_{i j} \frac{1}{4 f^2} (k_j^0 + k_i^0) \ ,
\end{equation}
where $k^{0}_{i,j}$ are the energies of the mesons
and the explicit values of the coefficients $C_{ij}$ can be
found in Ref. \cite{oset98}. 
The coupled-channel Bethe Salpeter equation with the kernel (potential) 
$V_{i j}$
was used in \cite{oset98} in order to obtain the elastic and transition matrix
elements in the $K^- N$ reactions. The diagrammatic expression of this series
can be seen in fig. 1.
\begin{figure}[htb]
 \begin{center}
\includegraphics[height=1.8cm,width=12.cm,angle=0] {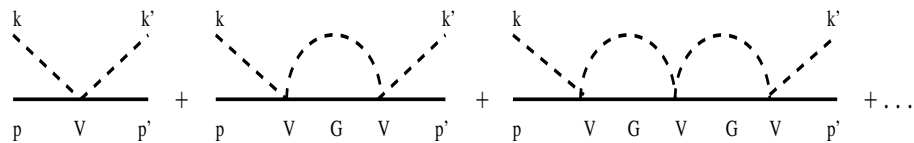}
 \caption{Diagrammatic representation of the Bethe-Salpeter equation.}
 \end{center}
\end{figure}
 The Bethe Salpeter
equations in the center of mass frame read
\begin{equation}
T_{i j} = V_{i j} + {V_{i l} \; G_l \; T_{l j}}  ,
\end{equation}
where the indices $i,l,j$ run over all possible channels and $G_l$ stands
for the loop function of a meson and a baryon propagators. Although in the
former equation the last term on the right hand side involves 
in principle
the off shell dependence of the amplitudes, it was shown in \cite{oset98} that
the amplitudes factorize on shell in the integral, the off shell part being
absorbed into a renormalization of the coupling constant $f_\pi$.

As shown in \cite{oset98} the results obtained lead to quite good agreement with
experimental data for cross sections of the different reactions and the mass
distribution of the $\Lambda(1405)$ resonance which is seen in the invariant
mass distribution of $\pi \Sigma$ produced in some reactions.

 One may wonder why the lowest order Lagrangian gives already good results while
in principle there should be important contributions from the next to leading
order Lagrangians.  This was also the case for the meson meson interaction in
the s-wave, as shown in \cite{oset97}. The reason can be seen in \cite{iam}
where a more general treatment of the problem was done in all channels of the
meson meson interaction. This method is the inverse amplitude method (IAM) in
coupled channels which uses the $L^2$ and $L^4$ chiral Lagrangians, imposes
unitarity exactly and makes a chiral expansion of the real part of the
inverse of the T matrix. The order $O(p^4)$ contribution to the T matrix
$T_4$ contains the loop function with two matrices at order $O(p^2)$, $T_2$,
and the loop function of two mesons, plus the polynomial
contribution $T_4^P$ which comes from the $L^4$ chiral Lagrangian.  The loop
function is divergent and must be regularized, with either dimensional
regularization, cut off, etc.  The total $T_4$ contribution sums these two
pieces and is independent of the cut off. This means that we can make a trade
off with the cut off and the $T_4^P$ contribution such as to minimize the
contribution of the latter. If then this contribution can be neglected,
 the IAM method leads to the Bethe Salpeter equation with $T_2$ playing the
 role of the potential. This is possible for the s-wave, both in meson meson
 and in the $\bar{K} N$ interaction, but can not be used for p-waves where
 explicit resonances \cite{nsd}, or alternatively the $L^4$ Lagrangians must be
 considered \cite{iam}.

 In order to evaluate
the $\bar{K}$ selfenergy, one needs first to include the medium modifications 
in the $\bar{K} N$ amplitude, $T_{\rm
eff}^{\alpha}$ ($\alpha={\bar K}p,{\bar K}n$), and then perform the
integral over the nucleons in the Fermi sea: 

\begin{equation}
\Pi^s_{\bar{K}}(q^0,{\vec q},\rho)=2\int \frac{d^3p}{(2\pi)^3}
n(\vec{p}) \left[ T_{\rm eff}^{\bar{K}
p}(P^0,\vec{P},\rho) +
T_{\rm eff}^{\bar{K} n}(P^0,\vec{P},\rho) \right] \ ,
\label{eq:selfka}
\end{equation}

The values
$(q^0,\vec{q}\,)$ stand now for the energy and momentum of the
$\bar{K}$ in the lab frame, $P^0=q^0+\varepsilon_N(\vec{p}\,)$,
$\vec{P}=\vec{q}+\vec{p}$ and $\rho$ is the nuclear matter density.

We also include a p-wave contribution to the ${\bar K}$ 
self-energy coming from the coupling of the ${\bar K}$ meson to
hyperon-nucleon hole ($YN^{-1}$) excitations,
with $Y=\Lambda,\Sigma,\Sigma^*(1385)$. The vertices $MBB^\prime$ are 
easily derived from
the $D$ and $F$ terms of Eq.~(1). The explicit expressions can be seen in
 \cite{knuc}. 
 At this point it is interesting to recall three different approaches to the
 question of the $\bar{K}$ selfenergy in the nuclear medium. The first
 interesting realization was the one in \cite{koch94,wkw96,waas97}, 
 where the Pauli blocking in the intermediate nucleon states in
 fig.1 induced as a results a shift of the $\Lambda(1405)$ resonance to higher
 energies and a subsequent attractive $\bar{K}$ selfenergy. The work of
 \cite{lutz} introduced a novel an interesting aspect, the selfconsistency.
 Pauli blocking required a higher energy to produce the resonance, but having a
 smaller kaon mass led to an opposite effect, and as a consequence the
 position of the resonance was brought back to the free position. Yet, a 
 moderate
 attraction on the kaons still resulted, but weaker than anticipated from the
 former work.  The work of \cite{knuc} introduces some novelties. It
 incorporates the selfconsistent treatment of the kaons done in \cite{lutz} 
 and in addition it also includes the selfenergy of the pions, which are let to
 excite ph and $\Delta h$ components. It also includes the  mean field
 potentials of the baryons.  
\begin{figure}[htb]
 \begin{center}
\includegraphics[height=8.cm,width=10.cm,angle=0] {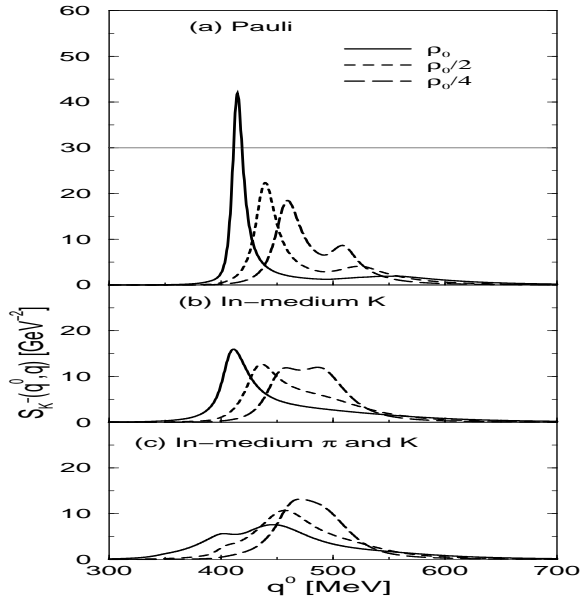}
 \caption{Kaon spectral function at several densities.}
 \end{center}
\end{figure}
The obvious consequence of the work of \cite{knuc}
is that the spectral function of the kaons  gets much wider than in the two
former approaches because one is including new decay channels for the
$\bar{K}$ in nuclei. This can be seen in fig. 2.
The work of \cite{knuc} leads to an attractive potential around nuclear matter
density and for kaons close to threshold of about 40 MeV and a width of 
about 100 MeV.

\section{ Kaonic atoms}

  In the work of \cite{zaki} the kaon selfenergy discussed above has been 
  used for the case of kaonic atoms, where there are
abundant data to test the theoretical predictions. One uses the Klein 
Gordon equation and obtains two families of states. One
family corresponds to the atomic states, some of which are those already  
measured, and 
which have  energies around or below 1 MeV and widths
of about a few hundred KeV or smaller. The other family corresponds to 
states which are nuclear deeply bound states, with
energies of 10 or more MeV and widths around 100 MeV. 
\begin{figure}[htb]
 \begin{center}
\includegraphics[height=7.cm,width=9.cm,angle=0] {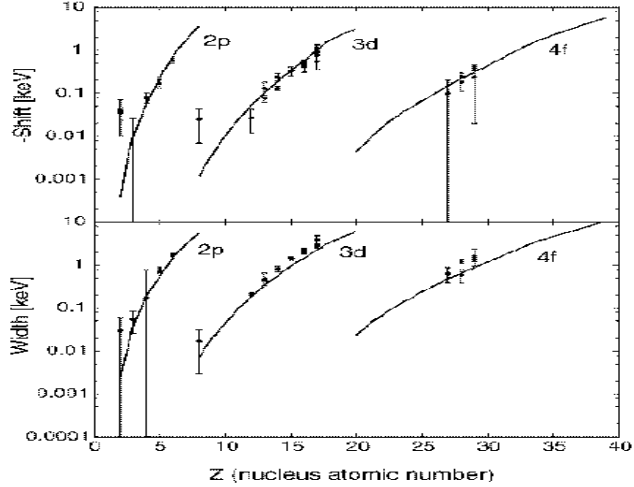}
 \caption{ Shifts and widths of kaonic atoms.}
 \end{center}
\end{figure}
 In fig. 3 we can see the results obtained for shifts and widths for 
 a large set of nuclei around the periodic table. The agreement
with data is sufficiently good to endorse the fairness of the theoretical 
potential. A best fit with a strength of the potential
slightly modified around the theoretical values can lead to even better
 agreement as shown in \cite{baca} and serves to quantify
the level of accuracy of the theoretical potential, which is set there at 
the level of 20-30 per cent as an average. The curious thing
is that there are good fits to the data using potentials with a strength 
at $\rho=\rho_0$ of the order of 200 MeV \cite{gal}. 
As shown in  \cite{zaki}, the results obtained there and those obtained using 
the potential
 of  \cite{gal} are in excellent agreement for
the atomic states. The differences in the two potentials 
appear in the deeply bound nuclear states. The deep potential
provides extra states bound by about 200 MeV, while the potential of 
\cite{knuc} binds states at most by 40 MeV.  This
remarkable finding can be interpreted as saying that the extra bound states, 
forcing the atomic states to be orthogonal to them, introduces
 extra nodes in the wave function and pushes the atomic states more
 to the surface of the nucleus, acting effectively as a
repulsion which counterbalances the extra attraction of the potential.  
This observation also tells that pure fits to the $K^-$
atoms are not sufficient to determine the strength of the $K^-$ nucleus 
potential. Other solutions with even more attraction at 
$\rho=\rho_0$ are in principle possible, provided they introduce new states 
of the deeply bound nuclear family.  On the other
hand, the work of \cite{baca} also tells us that at least an attraction as
 the one provided by the theoretical potential is needed. 

\section{ Scalar mesons} 

  Having the selfenergy of the kaons under control one can tackle new 
  problems where the kaon interaction with the medium is a
necessary ingredient.  One of such cases is the modification of the
 $f_0(980)$ and $a_0(980)$ resonances in the nuclear
medium. As shown in \cite{oset97} and \cite{iam} one can obtain a good 
description of the scalar resonances within the context
of the chiral unitary approach.  In particular, in \cite{oset97} it is 
possible to generate them using the lowest order Lagrangian and
the Bethe Salpeter equation, as also done for the $\bar{K}N$ interaction
in section 2. Here the coupled channels are
$K\bar{K}$ and $\pi\pi$ for the I=0 channel, where the $f_0$ resonance 
appears, and $K\bar{K}$  and $\pi\eta$ in the I=1
channel where the $a_0$ resonance appears.  In this case we must introduce 
in the Bethe Salpeter equation the proper
selfenergies of pions, kaons and eta, and in addition one has to 
include also vertex corrections which cancel the off shell part of
the vertices \cite{davesne,chiang}. 
\begin{figure}[htb]
 \begin{center}
\includegraphics[height=7.cm,width=9.cm,angle=0] {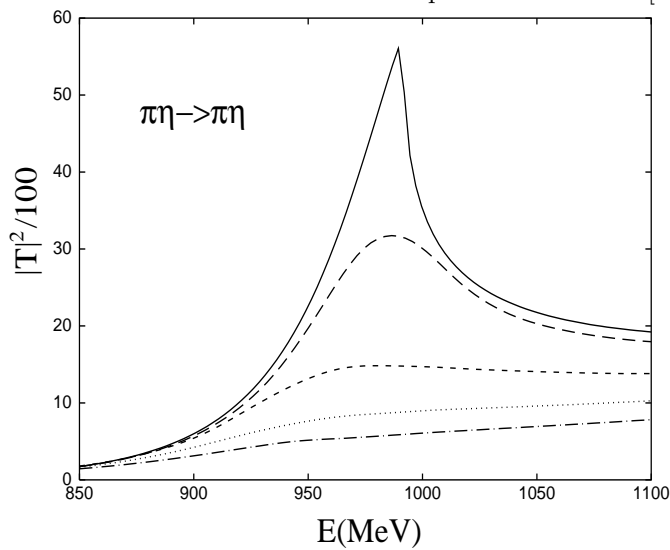}
 \caption{Modulus squared of the $\pi\eta$ scattering amplitude.
  Solid line, free amplitude; long dashed line, 
$\rho=\rho_0/8$; short dashed line, $\rho=\rho_0/2$; dotted line, 
$\rho=\rho_0$; dashed dotted line, $\rho=1.5\rho_0$.}
 \end{center}
\end{figure}
 The explicit calculations 
are done in \cite{manolo} and we show the results in
figs.  4 and 5 for different densities. 
\begin{figure}[htb]
 \begin{center}
\includegraphics[height=7.cm,width=9.cm,angle=0] {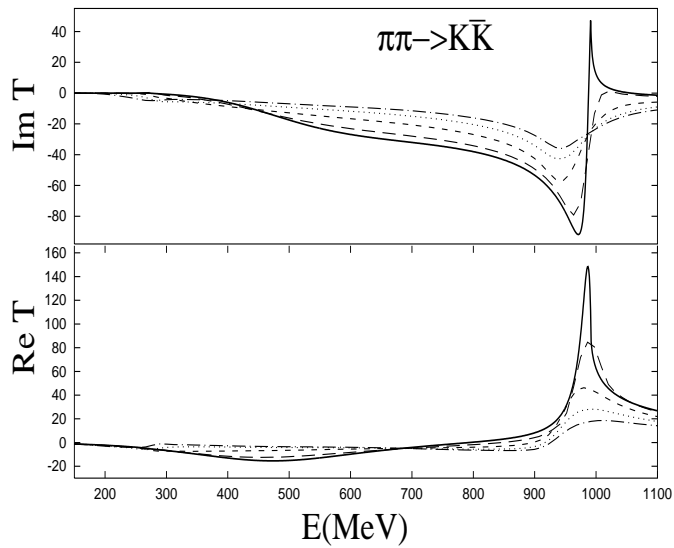}
 \caption{Imaginary and real part of the $\pi K$ scattering amplitude.
Lines have the same meaning as in the previous figure.}
 \end{center}
\end{figure}
 For the case of  I=1 we have 
shown the modulus squared of the  $\eta \pi$ amplitude, since
this is what would appear in the reaction where one looks at the invariant 
mass of the $\eta\pi$ system in the final state. What we
observe in the figures is that for the case of the $a_0$ resonance the 
shape becomes gradually wider as the density increases and
at densities of around $\rho_0$  the resonance is already 
washed away in the medium.
The case of the $f_0$ resonance, which is already narrower to begin with, 
is more hopeful, since even at $\rho_0$ one still can see
the resonance shape, however, the width passes from a free value around 
40 MeV to about more than 150 MeV. How to observe
these changes might not be so easy. In the talk of \cite{cassing} it was 
shown that using pions or protons as probes it was rather
difficult . Using photoproduction of two pions, and $\pi\eta$ photoproduction,
 might give better chances \cite{marco}. Other
possibilities would come from radiative decay of the $\phi$ in nuclei where 
recent experiments show a clean $f_0$ and $a_0$
peak \cite{exp1,exp2}, which incidentally can be very well described 
within the approach followed here to generate the
resonances \cite{uge} and a similar one where  a separable meson
meson potential is used rather than the amplitudes provided by the chiral 
Lagrangians \cite{markus}. 

\section{ $\phi$ decay in nuclei}

Finally let us say a few words about the $\phi$ decay in nuclei. The work 
reported here \cite{phi} follows closely the lines of
\cite{klingl,norbert}, however, it uses the updated $\bar{K}$ 
selfenergies of \cite{knuc}.  In the present
case the $\phi$ decays primarily in $K\bar{K}$, but these kaons can 
now interact with the medium as discussed previously.    For
the selfenergy of the $K$, since the $KN$ interaction is not too strong 
and there are no resonances, the $t\rho$ approximation is
sufficient.
\begin{figure}[htb]
 \begin{center}
\includegraphics[height=6.cm,width=8.cm,angle=0] {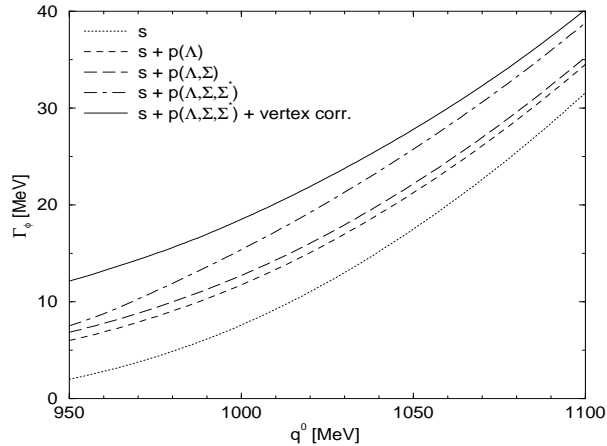}
 \caption{$\phi$ width at $\rho=\rho_0$.}
 \end{center}
\end{figure}
In fig. 6 we show the results for the $\phi$ width at $\rho=\rho_0$ 
as a function of the mass of the $\phi$, separating the
contribution from the different channels. What we observe is that the 
consideration of the s-wave $\bar{K}$-selfenergy is
responsible for a sizeable increase of the width in the medium, but 
the p-wave is also relevant, particularly the $\Lambda h$
excitation and the $\Sigma^*h$ excitation. It is also interesting to 
note that the vertex corrections \cite{beng}  (Yh loops attached
to the $\phi$ decay vertex)  are now present and do not cancel off 
shell contributions like in the case of the scalar mesons. Their
contribution is also shown in the figure and has about the same 
strength as the other p-wave contributions.  The total width of
the $\phi$ that we obtain is about 22 MeV at $\rho=\rho_0$, about 
a factor two smaller than the one obtained in
\cite{klingl,norbert}, yet, the important message is  the 
nearly one order of magnitude increase of the width with respect to
the free one. 
   We are hopeful that in the near future one can measure the width 
   of the $\phi$ in the medium, from
heavy ion reactions or particle nucleus interactions, although it will 
require careful analyses as shown in \cite{indio} for the case
of $K\bar{K}$ production in heavy ion collisions, where consideration 
of the possibility that the observed kaons come from
$\phi$ decay outside the nucleus leads to nuclear $\phi$ widths 
considerably larger than the directly observed ones. 

\section{ Summary} 

In summary we have reported here on recent work which involves the 
propagation of kaons in the nuclear medium. All them
together provide a test of consistency of the theoretical 
ideas and results previously developed and reported here. If we
gain confidence in those theoretical methods one can proceed to 
higher densities and investigate the possibility of kaon
condensates in neutron stars \cite{kaplan}.  The weak strength of our 
$\bar{K}$ potential  would make however the phenomenon highly
unlikely. 
  
On the other hand, we can also extract some conclusions concerning the 
general chiral framework: 
1) The chiral Lagrangians have much information in store. 
2) Chiral perturbation theory allows one to extract some of this 
information. 
3) The chiral unitary approach allows one to extract much more information. 
4) These unitary methods combined with the use of standard many body 
techniques are opening the door to the investigation of
new nuclear problems in a more accurate and systematic way, giving
 rise to a new field which could be rightly called "Chiral
Nuclear Physics". As chiral theory becomes gradually a more accepted
tool to deal with strong interactions at intermediate energies, chiral 
nuclear physics is bound to follow 
analogously in the interpretation of old and new phenomena in
nuclei.

\section*{Acknowledgments}

This work is partially supported
by DGICYT contract numbers PB98-1247 and PB96-0753,
 and by the
EEC-TMR Program under contract No. CT98-0169.

\end{document}